\documentclass[journal]{IEEEtran}
\usepackage{hyperref}
\usepackage{cite}
\usepackage[pdftex]{graphicx}
\usepackage[cmex10]{amsmath}
\usepackage{mdwmath}
\usepackage{array,booktabs,siunitx,xcolor,colortbl} 

\begin{document}

\title{Hybrid Plasmonic Waveguide Fed Broadband Nano-antenna for Nanophotonic Applications}

\author{Md~Saad-Bin-Alam, Md~Ibrahim~Khalil, Atiqur~Rahman,~\emph{Member,~IEEE,} and
Arshad~M.~Chowdhury,~\emph{Member,~IEEE}

\thanks{ Md~Saad-Bin-Alam, Md~Ibrahim~Khalil, Atiqur~Rahman and Arshad~M.~Chowdhury are with the Department of Electrical \& Computer Engineering, North South University, Dhaka-1229, Bangladesh. Md~Ibrahim~Khalil is also with School of Electrical and computer Engineering, Georgia Institute of Technology, Atlanta, GA-30309, USA}

\thanks{ Corresponding authors \href{ibrahim.khalil@ece.gatech.edu}{ibrahim.khalil@ece.gatech.edu}}

\thanks{\emph{\textcolor{blue}{First and Second author contributed equally throughout this work.}}}

\thanks{Color versions of one or more of the figures in this paper are available online
at \href{http://ieeexplore.ieee.org}{http://ieeexplore.ieee.org}}}

\markboth{}{}
\maketitle

\begin{abstract}
In this paper, we propose a novel hybrid plasmonic waveguide fed broadband optical patch nano-antenna for nanophotonic applications. Through full wave electromagnetic simulation, we demonstrated our proposed antenna to radiate and receive signal at all optical communication windows (e.g. $\lambda$ = 850nm, 1310nm  \& 1550nm) with around 86\% bandwidth within the operational domain. Moreover numerical results demonstrate that the proposed nano-antenna has directional radiation pattern with satisfactory gain over all three communication bands. Additionally, we evaluated the antenna performances with two different array arrangements (e.g. one dimensional and square array). The proposed broadband antenna can be used for prominent nanophotonic applications such as optical wireless communication in inter and intra-chip devices, optical sensing and optical energy harvesting etc.
\end{abstract}


\section{Introduction}
\PARstart{A}NTENNAS play an important role traditionally for many microwave applications. However, recently developments of antennas in the near-infrared and optical regions are showing great promises with better manipulation,  emission control and radiation of light waves into free space \cite{Novotly_Nature, Shegai_Scatter, Biosensing, photodetector, Heat_Transfer, Leily_on_Chip, Resonator, Energy_Harvesting}. In recent years, nano scale optical antennas are proposed for several applications in near-infrared, far-infrared and visible range. For example, nano-antennas have been considered as efficient and promising element in scattering \cite{Shegai_Scatter}, sensing \cite{Biosensing}, photo-detection \cite{photodetector}, heat-transfer \cite{Heat_Transfer}, inter- and/or intra-chip optical communications \cite{Leily_on_Chip, Resonator}, energy harvesting \cite{Energy_Harvesting} etc. Moreover, optical nano-antenna can play a vital role in reducing power consumption and allow higher speed for on-chip optical interconnects. As a result, it is now considered of as a prominent alternative to optical waveguide based interconnects for on-chip wireless communication that meets high spectral efficiency.

\begin{figure}[htb]
\centerline{\includegraphics[width=8.50cm]{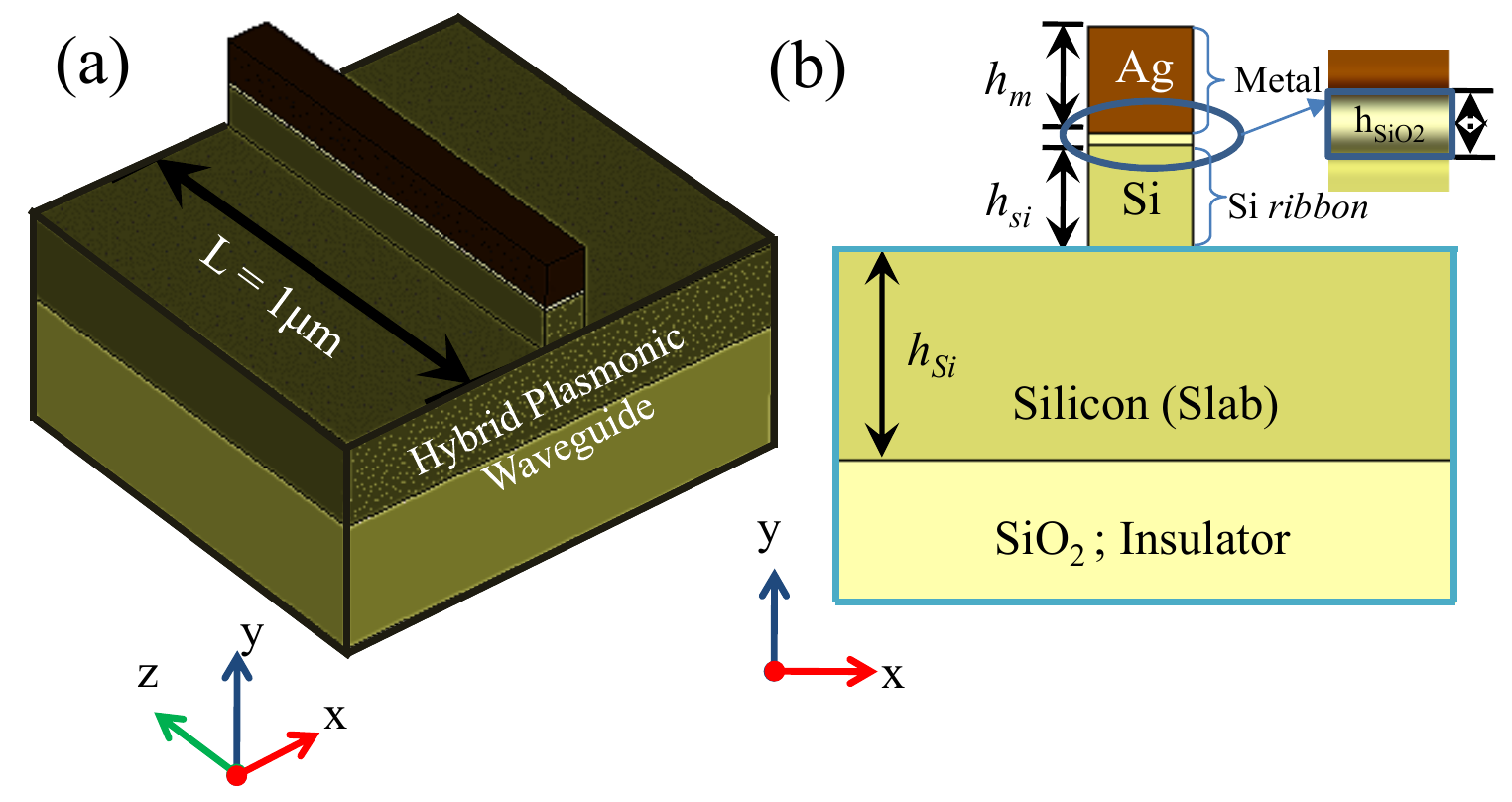}}
\caption{Schematic of 1$\mu$m long hybrid plasmonic waveguide where a layer of silicon di-oxide ($SiO_2$) with silver (Ag) cap embedded on a thin silicon (Si) layer (a) 3-D perspective view (b) cross-sectional view.}
\label{Figure1}
\end{figure}

The basic principle of designing an optical antenna is the optimization of shapes in order to match the impedance. Various shapes of optical antennas have been inspired from different designs of microwave antennas such as Rectangular and Circular Patch \cite{Leily_on_Chip,Resonator}, Dipole or Bow-tie \cite{bow-tie}, Yagi-Uda \cite{Yagi-Uda}, Spiral \cite{spiral}, Horn \cite{horn1,horn2,horn3}, J-pole and `\emph{Vee}' shaped \cite{J-pole}, Tapered-slot \cite{Energy_Harvesting}, Leaky-wave slot \cite{leaky-wave}, metallic Nano-slit with Cavity \cite{nano-slit} etc. However, simulation of optical antennas requires extra care since metals become dispersive in the visible region and need to be modeled with proper dielectric function. While the antenna proposed in \cite{Resonator} was implemented with dielectric resonator, in most cases nano-antennas operating at optical regime works on the principle of plasmonic resonance \cite{Novotly_Nature,bow-tie,J-pole}. Besides these techniques,  the concept of hybrid plasmonic structure has also been adopted for designing nano-antennas \cite{Leily_on_Chip,nano-slit}. In ref. \cite{Leily_on_Chip} authors have proposed a rectangular patch nano-antenna to radiate the localized light wave energy of a hybrid plasmonic waveguide \cite{plasmonic-waveguide,waveguides_bends,Gain_enhancement_Waveguide} where the signal of light was confined in ultra-thin low refractive index material in between a plasmonic metallic layer and a high refractive index material.

In previous design of modeling a nano-antenna the basic hybrid-plasmonic waveguide of ref. \cite{plasmonic-waveguide} was used to show that the antenna can radiate at 3\textsuperscript{rd} optical communication wavelength ($\lambda\approx$ 1550nm) \cite{Leily_on_Chip}. The main purpose of using the similar waveguide of Ref. \cite{Leily_on_Chip} is that it could confine light at the wavelength of 1550nm, although the confinement efficiency of the waveguide in other optical communication windows e.g. 1310nm and 850nm were not shown. Also traditional rectangular patch antenna is well-suited for narrow frequency bandwidth; hence the optical antenna modeled in \cite{Leily_on_Chip} was impedance-matched at 1550nm wavelength with 8\% bandwidth, which was sufficient for that particular wavelength.

\begin{figure}[htb]
\centerline{\includegraphics[width=8.50cm, height=7.8cm]{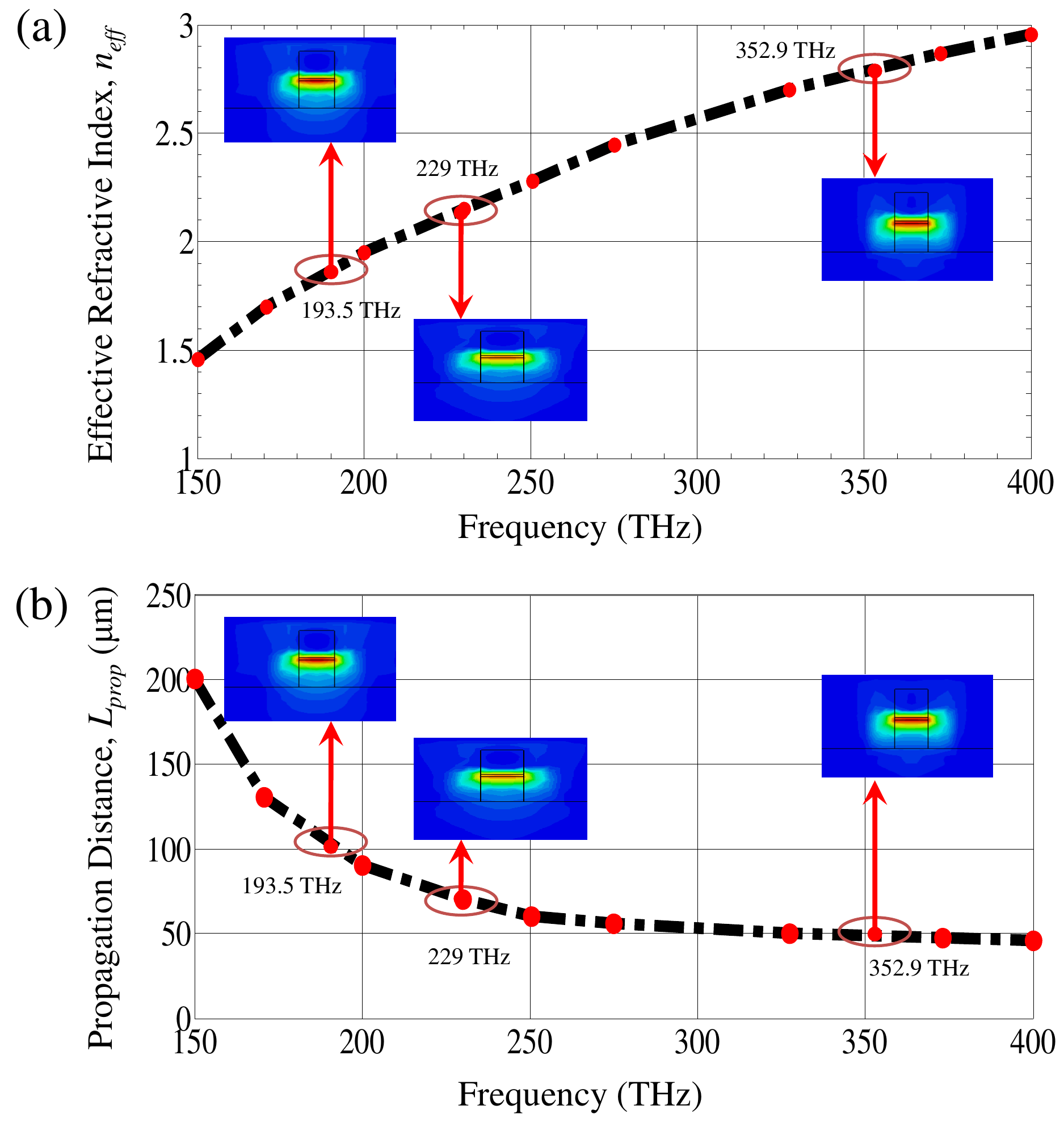}}
\caption{Effective refractive index and propagation distance of the hybrid plasmonic waveguide.}
\label{Figure2}
\end{figure}

In this paper, our objective is to study the performance of a hybrid plasmonic waveguide fed nano-antenna that can cover all optical communication wavelengths. Here we introduce a novel hybrid plasmonic ultra-wideband optical antenna which can be used with the waveguide mentioned above to radiate the localized energy or vice versa. Moreover we also observe the performance of our nano-antenna in array structure and found significant gain. This study confirms that proposed antenna can also have applications in energy harvesting.

\section{Characteristics of Hybrid Plasmonic Waveguide}
Before discussing the performance of our proposed nano-antenna, we will verify the efficiency of the hybrid plasmonic waveguide that will be connected to the proposed nano-antenna for three different optical communication windows such as 1550nm ($f_r$=193.5THz), 1310nm ($f_r$=229.0THz) and 850nm ($f_r$=352.9THz). Fig. \ref{Figure1}(a) provides a perspective view of 1$\mu$m long hybrid plasmonic waveguide. The cross-section of the waveguide along \emph{y}-axis is illustrated in Fig. \ref{Figure1}(b), where the thickness of metallic (Ag) layer, thin $SiO_2$ layer, $Si_{rib}$ layer and $Si_{slab}$ layer are defined as $h_m$ = 100nm, $h_{SiO_2}$ = 10nm, $h_{Si_{(rib)}}$ = 100nm and $h_{Si_{(slab)}}$ = 200nm. The width of the waveguide is considered as $W_{co}$ = 100nm. While the relative permittivity of $SiO_2$ is set to be $\epsilon_{SiO_2}$=2.09, that for Si is extracted from \cite{Optical_Silicon}. The relative permittivity of Ag is taken from \cite{Optical_Constants}. To verify the waveguide's efficiency, we carry out a two pronged approach: firstly we find out the effective refractive index, $n_{eff}$ of the hybrid plasmonic structure and later we simulate the propagation distance, $L_{prop}$ through the waveguide. These parameters have been calculated using CST mode solver \cite{CST_MWS}. We define propagation distance as the distance where the amplitude of the field decays to 1/e.  

\begin{figure}[htb]
\centerline{\includegraphics[width=7.50cm]{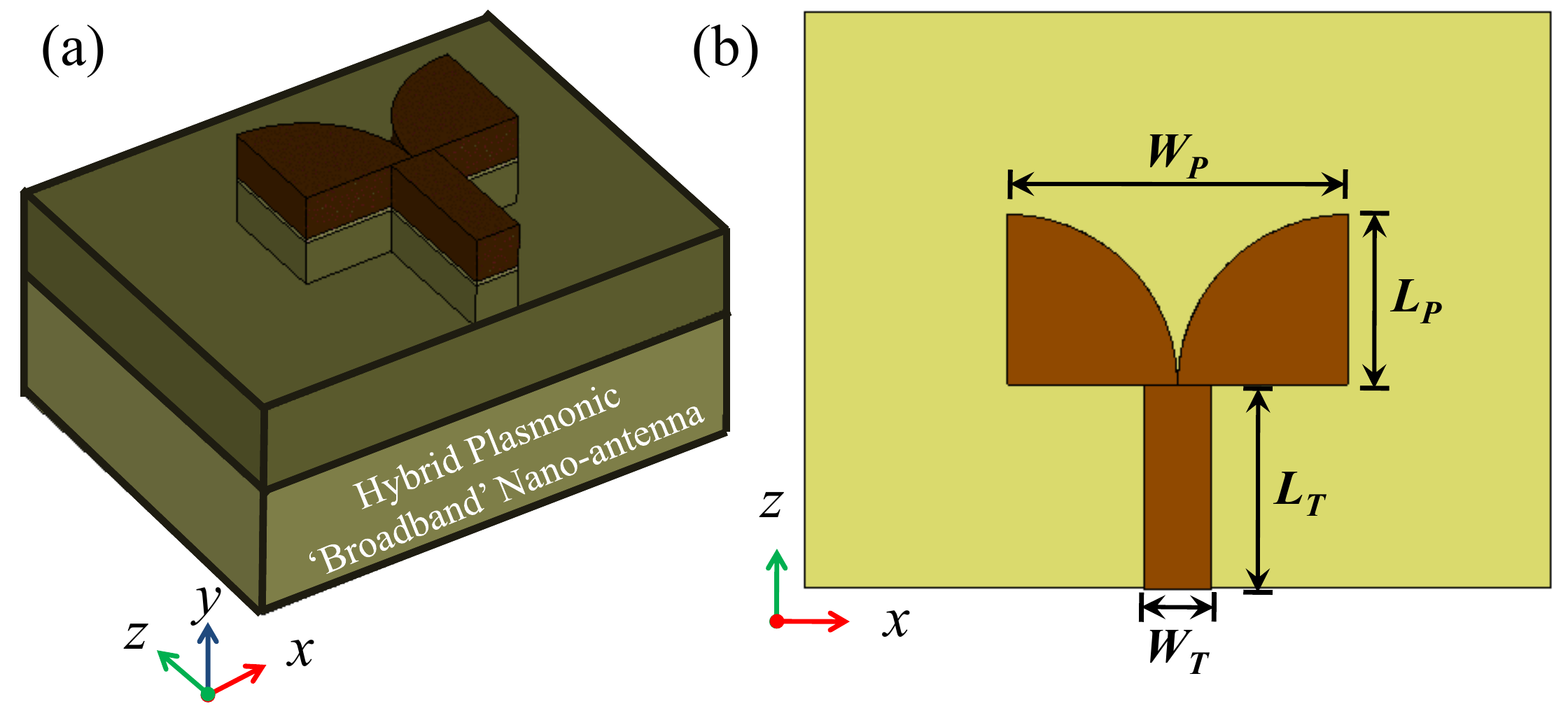}}
\centerline{\includegraphics[width=4.50cm]{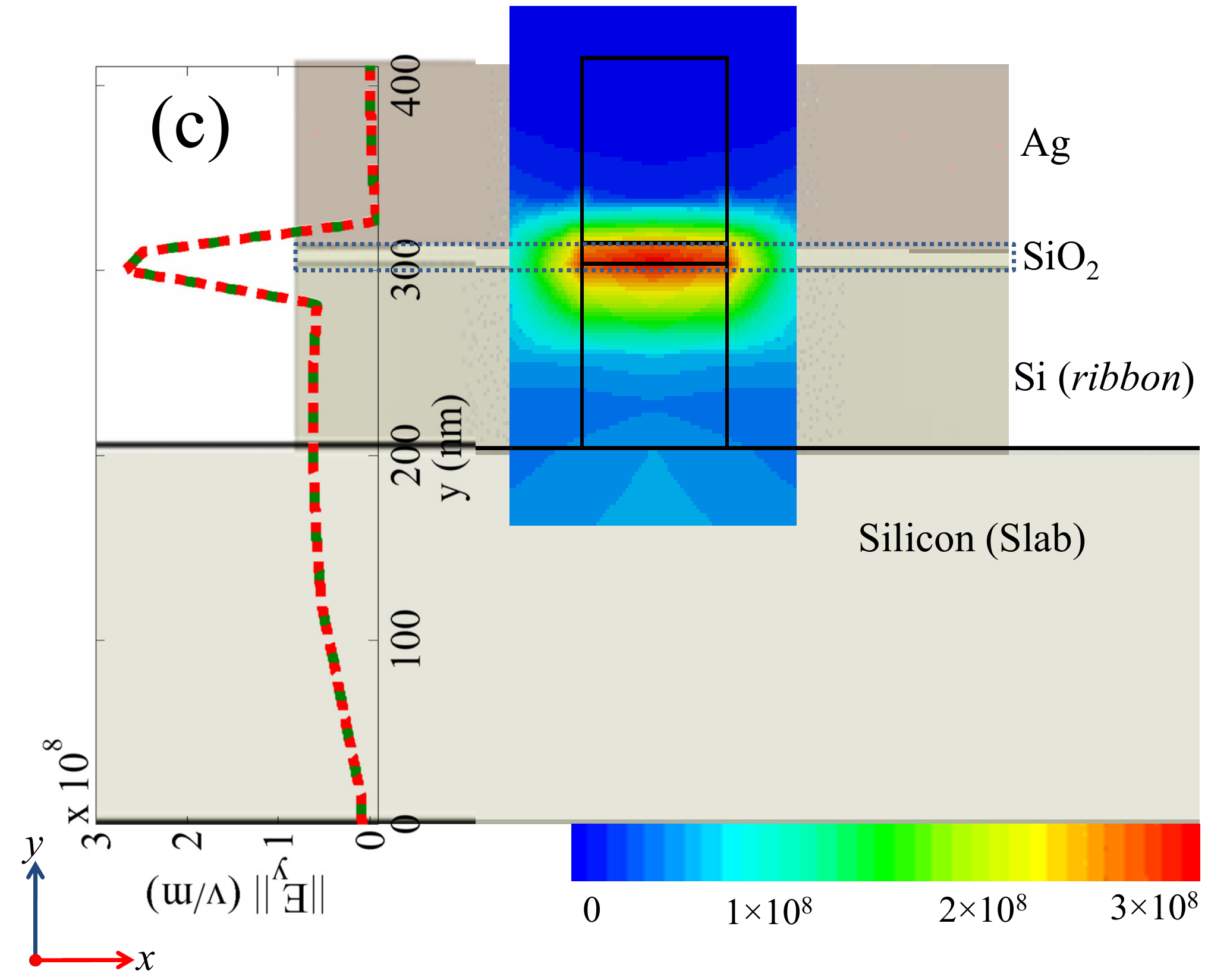}}

\caption{Schematic of the proposed hybrid plasmonic nano-antenna where a layer of silicon di-oxide ($SiO_2$) with silver (Ag) cap embedded on a thin silicon (Si) layer (a) 3-D perspective view (b) top view (c) 2-D cross-sectional view at the port side with $E$-field distribution at $z$ = 0.}
\label{Figure3}
\end{figure}

In step 1 we calculate the effective refractive index which is necessary to check the speed of light through the waveguide structure at different frequencies - as effective refractive index, $n_{eff}$ increases, the speed of light through the waveguide is decreased. Fig. \ref{Figure2}(a) reveals the value of $n_{eff}$ of the hybrid plasmonic waveguide at different frequencies of optical communication window (150THz--400THz) by keeping the dimensional parameters (thickness of the materials, width etc.) constant. It is clear from the curve that $n_{eff}$ of the waveguide increases gradually as frequency increases (wavelength decreases). This indicates the light speed is comparatively less in lower wavelengths than the wavelength of 1550nm (193.5THz). In the second step, the propagation distance, $L_{prop}$ ($\mu$m) is calculated against the frequency as shown in Fig. \ref{Figure2}(b); where we can see the propagation distance through the waveguide goes down with frequency which is due to the fact the losses become higher for shorter wavelength. For the three different optical communication frequencies, we can see the light (red portion) is successfully confined in the ultra-thin $SiO_2$ layer of the hybrid plasmonic waveguide structure. 

\section{Analytical Modeling of Nano-antenna}
To design a novel broadband optical nano-antenna including a transmission-line (which will be directly connected to waveguide), the aforementioned hybrid plasmonic structure is considered here. The nano-antenna can be fabricated using standard CMOS processing techniques which can be incorporated with the electronic circuit of an optical communication device. The perspective view of our designed optical nano-antenna is shown in Fig. \ref{Figure3}(a). Fig. \ref{Figure3}(b) shows the top view of our proposed nano-antenna. Visually this antenna is almost similar to the `\emph{Vivaldi Antenna}' often used in radio frequency range \cite{Vivaldi}. We numerically simulate the nanoantenna in CST\textsuperscript{TM} studio suite \cite{CST_MWS}. 

\begin{figure}[htb]
\centerline{\includegraphics[width=8.70cm]{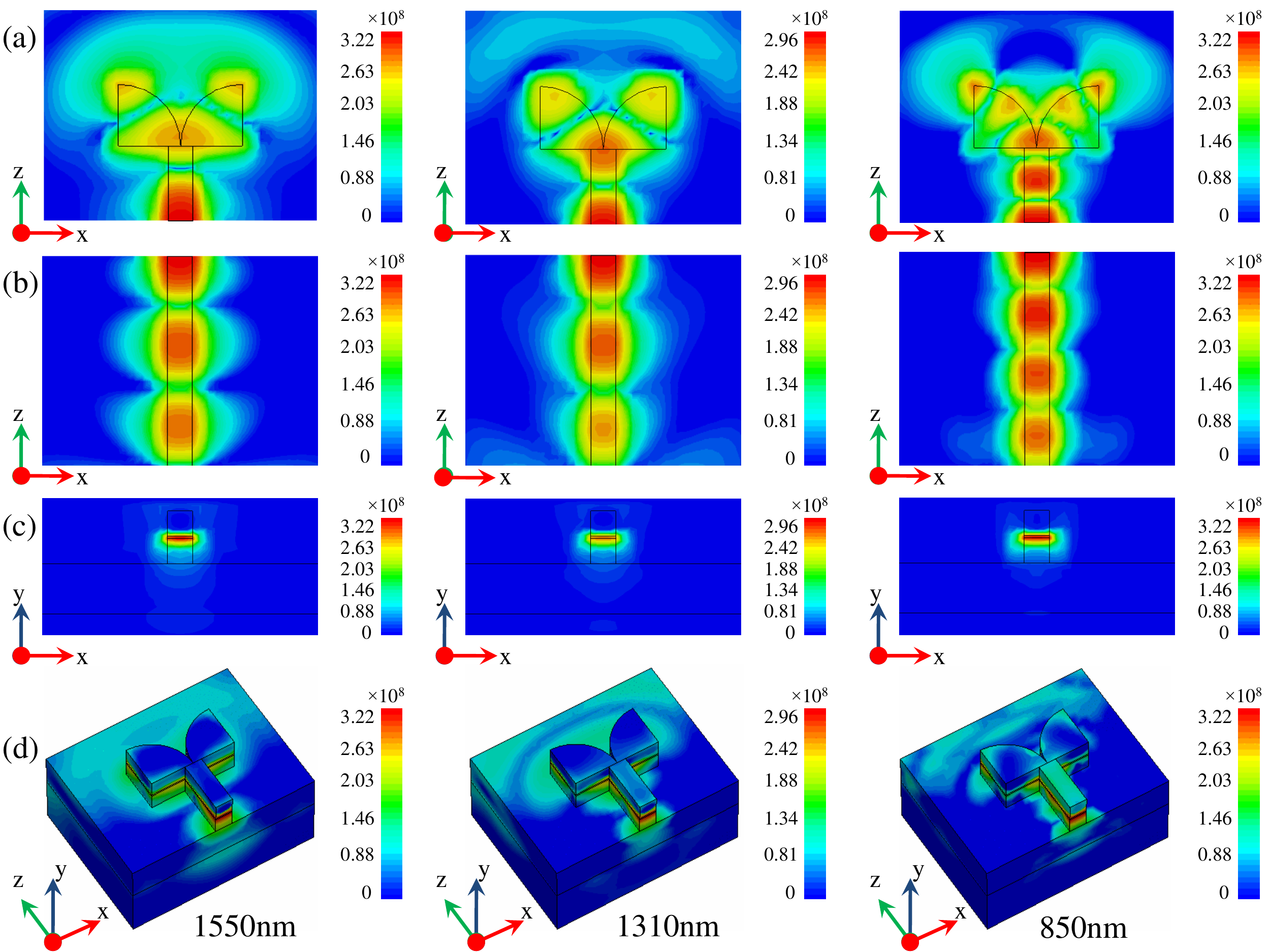}}
\caption{(a) E-field distributions of 2-D cross-section of the nano-antenna at $y$ = 305nm for $\lambda$ = 1550nm, $\lambda$ = 1310nm and $\lambda$ = 850nm; (b) $E$-field distributions of 2--D cross-section of the waveguide at $y$ = 305nm for $\lambda$ = 1550nm, $\lambda$ = 1310nm and $\lambda$ = 850nm; (c) $E$-field distributions of 2--D cross-section of the waveguide at $z$ = 0 for $\lambda$ = 1550nm, $\lambda$ = 1310nm and $\lambda$ = 850nm; (d) $E$-field distributions of 3--D perspective view of the nano-antenna for $\lambda$ = 1550nm, $\lambda$ = 1310nm and $\lambda$ = 850nm.}
\label{Figure4}
\end{figure}

To cover all optical communication windows, we fix our target frequency range of operation as 150THz-390THz ($\lambda$ = 2000nm-769.2nm) with the center frequency at $f_c$ = 270THz (\textit{$\lambda _c$} = 1111.1nm). We set our antenna's overall length, \textit{$L$ = $L_P$ + $L_T$} = 250nm + 300nm = 550nm; which is almost equal to half center wavelength, \textit{$\lambda _c$}/2 = 1111.1nm/2 = 555.55nm. The width is set to $W_P$ = 500nm. Each tapered-slot is designed as quarter-portion of a circle considering each lower-side corner of the nano-antenna as center and $L_P$ = 250nm is estimated as the radius of each circle. The width of transmission-line, $W_T$ remains same as the width of the hybrid plasmonic waveguide described above ($W_T$ = $W_{co}$ = 100nm). The overall dimension of the Si slab and its underneath $SiO_2$ insulator substrate is 1100nm $\times$ 850nm. The  port-side view of the nano-antenna is shown in Fig. \ref{Figure3}(c) which reveals that the $y$-component of electric field, $|E_y|$ of the port reaches the maximum amplitude which is confined in the thin $SiO_2$ layer as predicted in Fig. \ref{Figure2}. 

The electric field (or near-field) distributions at the wavelengths of 1550nm, 1310nm and 850nm for the simulated hybrid plasmonic optical nano-antenna are presented in Fig. \ref{Figure4}(a) showing how the optical signal is released and radiated to free-space after entering to the nano-antenna following its confinement and travel through the hybrid plasmonic waveguide. The electric field maps of the 1$\mu$m long hybrid plasmonic waveguide (at $y$ = 305nm) for the corresponding wavelengths are demonstrated in Fig. \ref{Figure4}(b) where the hybrid plasmonic waveguide has been exaggerated in order to show of better view of field distribution. Fig. \ref{Figure4}(c) shows the cross-sections of the hybrid plasmonic waveguide (as well as transmission line of the nano-antenna) at $z$ = 0 for the related wavelengths to show the light confinement capability of the waveguide at the analogous wavelengths. Finally, Fig. \ref{Figure4}(d) presents the 3--D structures of the complete broadband nano-antenna for the consequent wavelengths.

\begin{figure}[htb]
\centerline{\includegraphics[width=8.20cm, height=8.0cm]{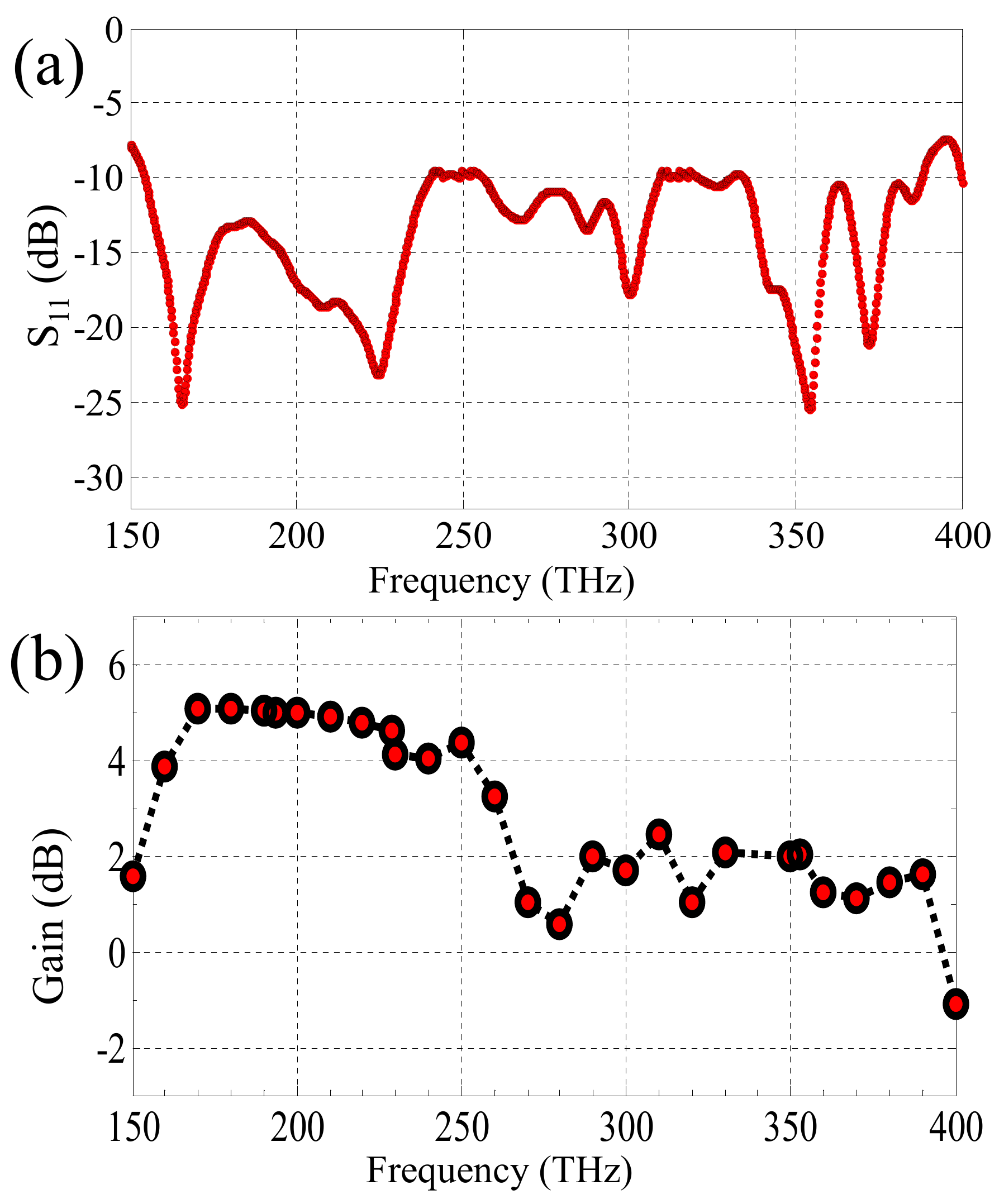}}
\caption{(a) Reflection coefficient and (b) gain of the proposed `\emph{Broadband}' nano-antenna.}
\label{Figure5}
\end{figure}

\begin{figure}
\centerline{\includegraphics[width=8.70cm]{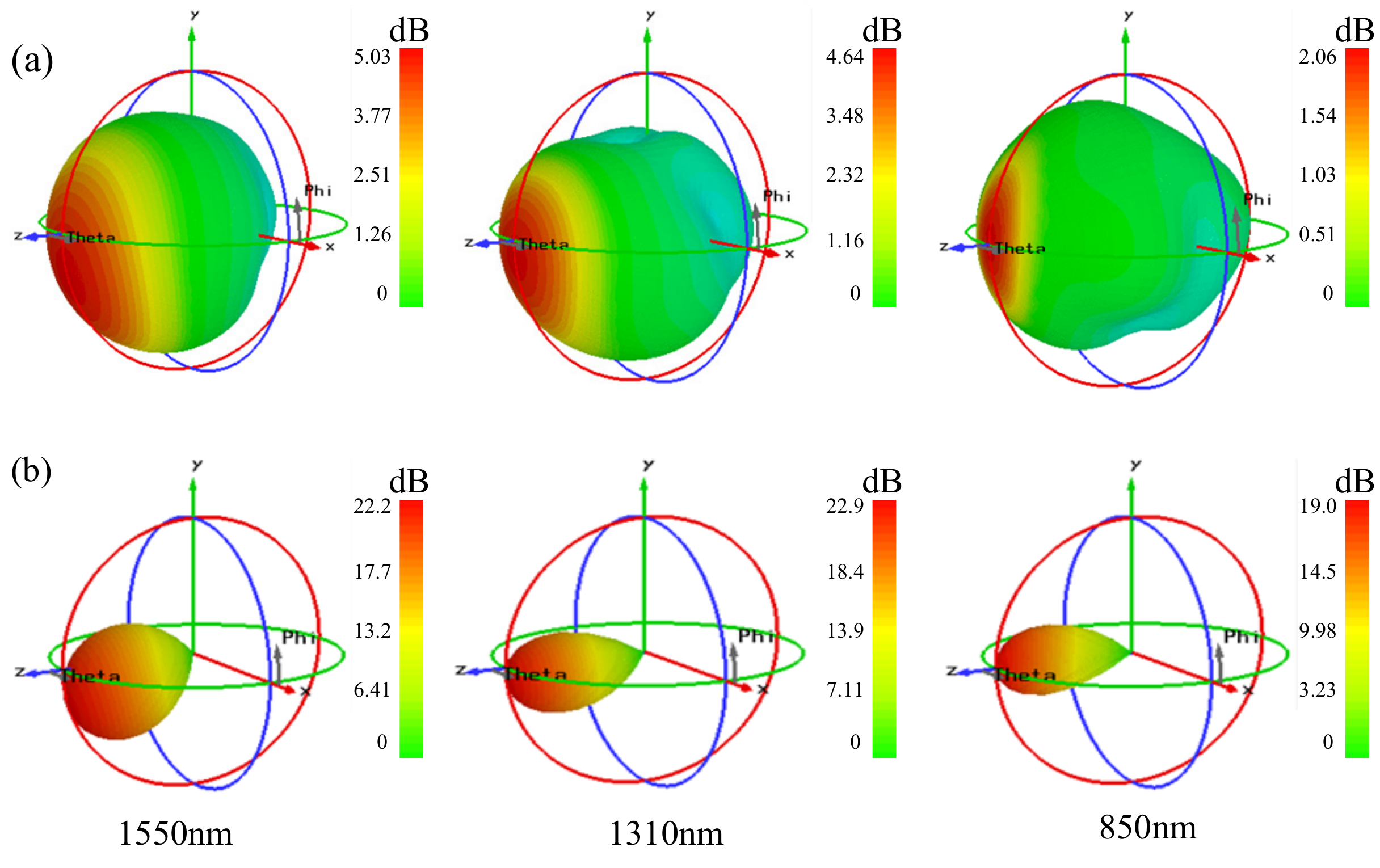}}
\caption{3-D radiation patterns of the proposed `\emph{Broadband}' nano-antenna for the wavelengths of 1550nm, 1310nm and 850nm (a) single nano-antenna (b) nano-antenna array. Number of elements is 100 for each type of array.}
\label{Figure6}
\end{figure}

The graph of the reflection coefficient ($S_{11}\leq$ -10 dB) shown in Fig. \ref{Figure5}(a) tells us that our simulated `\emph{Broadband}' nano-antenna can radiate optical signals to free-space and through reciprocity, receive optical signal from free space at the frequency range of 154THz to 388THz with a bandwidth of around 86\%, which covers our desired three optical communication wavelengths, in contrast to only 8\% bandwidth of the antenna proposed in \cite{Leily_on_Chip}. Here, we numerically calculate the center frequency, $f_c$ = 271THz ($\lambda_c$ = 1107nm). The gain of the antenna plotted against frequency in Fig. \ref{Figure5}(b) shows satisfactory values for any broadband/ultra wide band antenna \cite{horn2,horn3}. As revealed in Fig. \ref{Figure6}(a), the radiation patterns of our proposed antenna are directional to positive $z$-axis with the gain of 5.03dB, 4.64dB \& 2.06dB for the wavelengths of 1550nm, 1310nm \& 850nm respectively.

\begin{figure}
\centerline{\includegraphics[width=8.70cm]{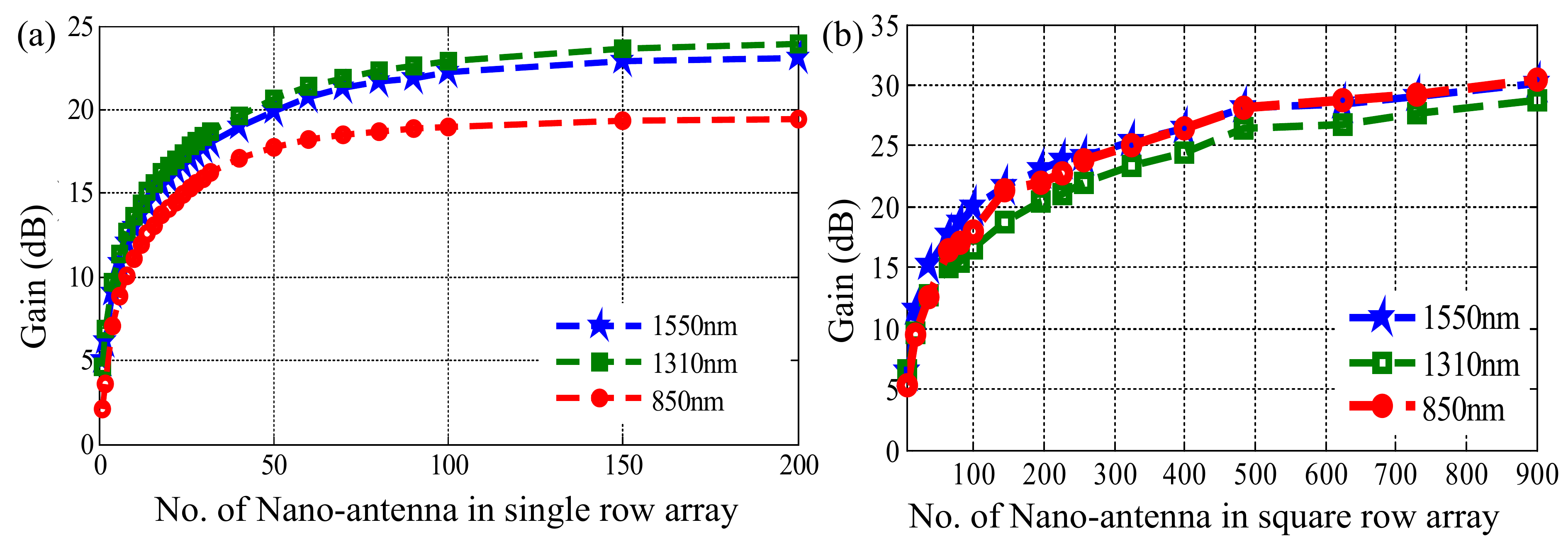}}
\caption{Gain against no. of `\emph{Broadband}' nano-antenna in (a) single row array and (b) square array for the wavelengths of 1550nm, 1310nm and 850nm.}
\label{Figure7}
\end{figure}

We also investigate  the performance of the nano-antenna when employed in an array. Such an investigation is useful if we want to apply our proposed antenna in optical energy harvesting applications like a nanoscopic rectifying antenna, or simply called `\emph{Nantenna}' \cite{Nantenna}, where a desirable level of gain and the radiation pattern of the array is required. To harvest a large amount of energy, the antenna must have a high gain; which can be obtained by creating an array of the nano-antenna. The information about the direction of radiation will help us to have the antenna orientation right with respect to the source signal. We examined two different types of array: first one is single row array and another is square array. We set the gap between two adjacent antennas as $250nm > \lambda_{eff}$; where $\lambda_{eff} = \frac{\lambda_c}{2}/n_{eff} = 230.6nm$; and for square array, the gap between two adjacent rows of antennas (added to $z$-direction) is set as $130nm > \lambda_{eff}$; where $\lambda_{eff} = \frac{\lambda_c}{4}/n_{eff} = 115.3nm$. The value of $n_{eff}$ = 2.4 is numerically extracted for $\lambda _c$ = 1107nm from Fig. \ref{Figure2}(a). In Fig. \ref{Figure7}(a,b), we can see the gain is gradually increasing as the number of the antenna is increased in each of the single row and square array for all optical communication windows, which is a desirable behavior of array. However, the gain of the array becomes saturated when the number of elements exceeds 100 for single row array and 500 for square array, which can be attributed to the increased coupling between elements when the number of elements is very high. Fig. \ref{Figure6}(b) shows the 3-D far-field radiation patterns for nano-antenna array at three optical communication wavelengths; which indicates the array of the proposed broadband nano-antenna (here, no. of antenna is 100$\times$1 = 100 for single row array, or 10$\times$10 = 100 for square array respectively) can radiate and receive signal directionally a gain nearly closed to or above 20dB for all optical communication windows.

\section{Conclusion}
In summary, our proposed hybrid plasmonic waveguide fed Broadband optical nano-antenna has the novelty in terms of the range of operational bandwidth as well as its compatibility with hybrid plasmonic optical waveguide that can carry the light signal corresponding to all optical communication windows - 1550nm, 1310nm and 850nm wavelengths. We have designed this unique shaped antenna (also known as `\emph{Vivaldi Antenna}'), which is not only valid for the microwave or radio frequencies, but also suitable in infrared/optical frequency domain. The results from full wave electromagnetic simulation confirmed that the proposed nano-antenna has directional radiation pattern with satisfactory gain over all three communication bands. Our proposed nano-antenna can be employed for various nano-photonic applications, in particular, for inter and intra-chip optical communications to establish wireless-optical link between optical circuits and would be very useful as it can cover all the optical communication wavelength. Besides, the proposed antenna can be used for optical energy harvesting (also known as nano-rectenna or \emph{Nantenna}), optical sensing because of its wide frequency coverage.

\end{document}